\documentclass[final,5p,times,twocolumn]{elsarticle}

\usepackage{graphicx}
\usepackage{amsmath}
\usepackage{color}

\journal{Physics Letters A}

\begin{document}

\begin{frontmatter}

\title{Ground-state correlation energy of counterions at a charged planar wall: Gibbs-Bogoliubov lower-bound approach
}
\author{Hiroshi Frusawa}

\ead{frusawa.hiroshi@kochi-tech.ac.jp}
\address{Laboratory of Statistical Physics, Kochi University of Technology, Tosa-Yamada, Kochi 782-8502, Japan}

\begin{abstract}
Recent simulation results imply the lowering of the ground-state correlation energy per counterion at a charged planar wall, compared with that of the 2D and 3D one-component plasma systems. 
Our aim is to correctly evaluate the ground-state energy of strongly-coupled counterion systems by considering a quasi-2D bound state where bound counterions are confined to a layer of molecular thickness. 
We use a variational approach based on the Gibbs-Bogoliubov inequality for the lower-bound free energy so that the liquid-state theory can be incorporated into the formulations. 
The soft mean spherical approximation demonstrates that the lowered ground-state energy can be reproduced by the obtained analytical form of a quasi-2D bound state.
\end{abstract}

\end{frontmatter}

\section{Introduction}
Macroions, including a macroscopic glass plate as well as colloids, are likely to carry a total surface charge exceeding thousands of elementary charges due to the release of counterions, mobile ions with opposite charges than the surfaces.
The macroion-counterion suspensions are characterized by high asymmetries between counterions and macroions in the valence of charges and by size, which contrasts with the properties of conventional electrolytes, such as salty water [1-5].
Because of high asymmetries, the macroion suspensions are necessarily inhomogeneous liquids, in that some counterions are electrostatically bound around macroion surfaces and form ionic clusters [1-5].

Our concern in this study is the ground state of such counterion systems in the strong coupling (SC) limit that can be achieved by either highly charged surfaces or counterions of high valency in low dielectric media, even at room temperature [2-15]. 
In the SC limit, where the counterion--counterion interaction strength is extremely large, a major part of the counterions exists in the proximity of the macroion surface [2-15], similar to the one-component plasma (OCP) filled with an electrically neutralizing background [2, 3, 16-25].
However, there is a crucial difference between counterion systems and the OCP due to the localization of macroion or the electrically neutralizing background.
In the first instance, the whole space in the OCP is filled with a smooth neutralizing background in either the 2D or 3D system [2, 16-25].
Conversely, counterions are spread over a 3D electric double layer that leads to violation of global electrical neutrality even in the ground state, while forming the quasi-2D Wigner crystalline layer of molecular thickness on the macroion surface [2-15].

Extensive Monte-Carlo simulations of the counterion systems in the SC regime have been performed [4, 5, 11-15].
In particular, the simulation studies have investigated the longitudinal density distribution along the vertical $z_0$-axis to the charged planar wall in one- and two-planar wall systems and verified the asymptotic behavior, herein referred to as the ground-state density distribution [2-15]. 
Among a variety of simulation studies, attention should be paid to a recent thorough investigation [15] of large-$z_0$ dependences. 
From these results [15], it is ubiquitously found in the SC regime that crossovers occur from the ground-state distribution in the vicinity of the wall to the mean-field behavior far from the wall;
the latter satisfies the Poisson-Boltzmann solution expressed by the effective Gouy--Chapman (GC) length, a characteristic length of the electric double layer [3, 6, 7, 15].

In this study, we first clarify theoretical issues on the ground-state correlation energy implied by the recent simulation results [15] on the above crossover behaviors. 
Specifically, the effective GC length [3, 6, 7, 15] determined by recent results reveal that the ground-state correlation energy per counterion, $u_{\infty}$, should be lower than the ground-state energies of the 2D and 3D OCP systems. 
We aim to resolve this discrepancy between the conventional OCPs and counterion systems by evaluating $u_{\infty}$ per counterion bound on the oppositely charged surface. 
Liquid-state theory [1], which is incorporated into the formulations via the Gibbs-Bogoliubov inequality [1, 25, 26], forms the basis of the present evaluation.

In Section 2, the $z_0$-dependence of longitudinal density distribution is summarized in Fig. 1; moreover, a combination of the above simulation results [15] and a previous theoretical model [3, 6, 7] reveals the above theoretical issues on the ground-state correlation energy $u_{\infty}$ per counterion.
Section 3 addresses the discrepancy of $u_{\infty}$ from the conventional 2D OCP results [16-25] by evaluating $u_{\infty}$ based on the liquid-state theory [1]. 
Accordingly, we can provide the correct evaluation of the ground-state energy by considering bound counterions that are confined to a quasi-2D layer of molecular thickness, consistently with the previous model used [3, 6, 7].
Section 4 validates the free energy functional for the above evaluation not only via the Gibbs-Bogoliubov inequality [1, 25, 26] but also via the density functional integral representation [26] that describes density fluctuations around the ground-state distribution. 
Final remarks are given in Section 5.

\begin{figure}[hbtp]
\begin{center}
	\includegraphics[
	width=7.7 cm
]{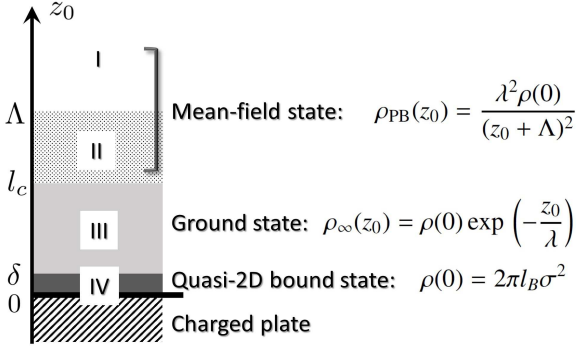}
\end{center}
\caption{Simulation results [15] on the longitudinal density distribution, which can be divided into four states. The first state (I. Mean-field state) is observed for $z_0\geq\Lambda$, being far from the charged planar wall where the density distribution can be described by the mean-field form, i.e. the same distribution as that of the Poisson--Boltzmann solution (eq. (\ref{PB})) with the Gouy--Chapman (GC) length $\lambda$ replaced by an effective GC length $\Lambda$ [3, 6, 7, 15]. We can also observe a density plateau for $l_c\leq z_0\leq \Lambda$ (II. Mean-field state) where the density ceases to increase despite approaching the planar wall due to the Poisson--Boltzmann solution. Region II is followed by the characteristic behavior of strongly coupled counterion systems for $z_0\leq l_c$ (III. Ground state) where the density distribution is governed by the external potential $z_0/\lambda$ created by the charged wall. In actuality, the ground state includes bound counterions that are confined to a layer of molecular thickness $\delta$, or region IV on the planar surface, which is referred to as a quasi-2D bound state (see section 3.3 for the definition of this state).　
Correspondingly, the contact density $\rho(0)=2\pi l_B\sigma^2$ is regarded as a coarse-grained density when considering the chemical equilibrium between II and IV states as given by eq. (\ref{chemical equilibrium}).
}
\end{figure}
\section{Ground-state correlation energy $u_{\infty}$ per counterion: theoretical issues}
\subsection{Recent simulation results [15] on the longitudinal density distributions}
We consider the counterion density distribution along the $z_0$-axis vertical to the charged planar wall having the charge density of $\sigma e$ per unit area, which will be referred to as the longitudinal distribution compared with the transverse one that is parallel to the charged planar wall. 
Recent Monte Carlo simulation results [15] on the longitudinal density distributions $\rho(z_0)$ have revealed the large-$z_0$ behaviors in a high range of the coupling constant $\Gamma$ such that $5\times10^3\leq\Gamma\leq2\times10^6$, where $\Gamma$ is defined as $\Gamma=q^2l_B/a$ using the counterion valence $q$, the Bjerrum length $l_B$ (the length at which two elementary charges interact electrostatically with thermal energy $k_BT$), and the Wigner-Seitz radius $a$ that satisfies the following relation [2-7]:
\begin{equation}
\pi a^2\sigma=q.
\label{ws}
\end{equation}
Equation (\ref{ws}) implies that a single counterion carrying charge with the absolute value of $qe$ neutralizes the surface charge over the area $\pi a^2$ of the charged planar wall.
Accordingly, the local electrical neutrality requires that the characteristic length $a$ be selected as the separation distance between counterions bound in the proximity of the charged planar wall, despite the violation of global electrical neutrality, i.e., the non-vanishing of total effective charges of the charged planar wall plus bound counterions.

Figure 1 summarizes the results [15], indicating that the $z_0$-dependences of $\rho(z_0)$ consist of four parts divided by three characteristic lengths: a molecular scale of ionic size $\delta$, a crossover distance $l_c$, and the effective GC length $\Lambda$ [3, 6, 7, 15].

Regions I and II, far from the planar wall in Figure 1, can be regarded as being in the mean-field state because the density profile in these regions obey the following solution of the Poisson-Boltzmann type [3, 6, 7, 15]:
\begin{equation}
\rho_{\mathrm{PB}}(z_0)=\frac{\lambda^2\rho(0)}{(z_0+\Lambda)^2},
\label{PB}
\end{equation}
using $\Lambda$, in addition to the original GC length $\lambda=1/(2\pi ql_B\sigma)$ and the contact density $\rho(0)$ at the charged planar wall.
The planar contact value theorem [3-5] uniquely determines that
\begin{equation}
\rho(0)=2\pi l_B\sigma^2,
\label{contact}
\end{equation}
irrespective of $\Gamma$.
Equation (\ref{PB}) converges to the contact density (\ref{contact}) at $z_0=0$ if $\Lambda=\lambda$. 
In actuality, the effective GC length $\Lambda$ is much larger than $\lambda$, and correspondingly $\rho_{\mathrm{PB}}(0)/\rho(0)\ll 1$ as described below.

The simulation results [15] demonstrate that these behaviors are precisely described by the above mean-field form (\ref{PB}) when selecting an appropriate length of $\Lambda$.
The details are as follows. 
For $\Gamma\geq 5\times10^3$, we can see that the large-$z_0$ density profile exhibits an algebraic decay ($z_0^{-2}$) for $z_0\geq\Lambda$ in region I of Fig. 1, which is connected to a plateau, a quasi-constant density, in the range of $l_c\leq z_0\leq\Lambda$ (the region II in Fig. 1). 
The mean-field surface has a density of
\begin{equation}
\rho_{\mathrm{PB}}(0)=\left(
\frac{\lambda}{\Lambda}
\right)^2\rho(0),
\label{mean surface}
\end{equation}
which is smaller than the above contact density $\rho(0)$ given by eq. (\ref{contact}).
The density plateau is retained to the mean-field surface on the II-III boundary in the range of $l_c\leq z_0\leq\Lambda$, which implies that region II corresponds to the electric double layer of the GC type in terms of the mean-field picture.

The density distribution deviates considerably from the mean-field distribution (\ref{PB}) when entering region III. 
The SC theory has shown in a field-theoretic manner [3, 4, 9, 10] that the ground-state density distribution $\rho_{\infty}(z_0)$ is given by
   \begin{equation}
   \rho_{\infty}(z_0)=\rho(0)\exp\,\left(-\frac{z_0}{\lambda}\right),
   \label{ground-density}
   \end{equation} 
which reads in the rescaled system of ${\bf r}=(x,y,z)\equiv{\bf r}_0/a$:
\begin{align}
\rho_{\infty}(z)&=a^3\rho(0)\,\exp\left(-2\Gamma z\right),\nonumber\\
a^3\rho(0)&=\frac{a}{\pi\lambda}=\frac{2\Gamma}{\pi},
\label{rescaled ground density}
\end{align}
where the relation $a/\lambda=2\Gamma$ was used.
Several Monte Carlo simulation results [4, 5, 11-15] have confirmed the above ground-state distribution $ \rho_{\infty}(z_0)$ in the vicinity of the planar surface (i.e., $0\leq z_0\leq l_c$ in Fig. 1), which is the reason why region III in Fig. 1 is called the ground state.

Equation (\ref{rescaled ground density}) is reduced to $\rho_{\infty}(z)=a^3\rho(0)\delta(z)$ in the limit of $\Gamma\rightarrow\infty$, implying that we can regard all counterions as bound ones at the charged planar wall in the SC limit of a coarse-grained system that can neglect the Gouy-Chapman length represented as $\lambda/a=(2\Gamma)^{-1}\rightarrow 0$ in the rescaled system.
While the ideal ground state of counterions has been identified with the 2D Wigner crystal in some previous models [3--7, 11, 12, 15], a magnified view leads to supposition that the bound state has a finite thickness of molecular scale $\delta$ as indicated in region IV of Fig. 1, considering that the bound counterions and oppositely charged planar wall are unable to merge.
In section 3.3, the quasi-2D bound state will be formulated using a cylindrical model. 

\subsection{Evaluation of $u_{\infty}$ from combining the simulation results and two-phase model [3, 6, 7]}
Let $u_{\infty}$ denote the correlation energy in the $k_BT$-unit per counterion in the ground state;
incidentally, the other energies and potentials used in this study are also defined by the $k_BT$-unit. 
We can write $u_{\infty}$ as
\begin{equation}
u_{\infty}=-\alpha\Gamma=-\alpha q^2l_B\left(\frac{\pi\sigma}{q}\right)^{1/2},
\label{correlation}
\end{equation}
with $\alpha$ being the prefactor specified below. 
In the second equality of eq. (\ref{correlation}), we have used relation (\ref{ws}).

The ground-state energy $u_{\infty}$ can be evaluated from combining two results: 
(i) the above simulation results [15] on the $\Gamma$-dependence of $\Lambda$, and (ii) the two-phase model [3, 6, 7] developed by Perel and Shklovskii (PS) [6], which provides the relationship between $\Lambda$ and the ground-state chemical potential $\mu_{\infty}$.

On the mean-field surface, or the boundary between regions II and III, an effective contact density $\rho_{\mathrm{PB}}(0)$ is given by eq. (\ref{mean surface}), whereas the actual contact density, $\rho(0)$ given by eq. (\ref{contact}), is much higher than that at the II-III boundary, $\rho_{\mathrm{PB}}(0)$, owing to the gain in the chemical potential $\mu_{\infty}$ associated with counterion-counterion correlations. 
The PS two-phase model [3, 6, 7] explains such a large difference of densities by considering the chemical equilibrium between the two regions, II and IV, in Fig. 1. We simply have
\begin{equation}
\ln\rho_{\mathrm{PB}}(0)=\ln\rho(0)+\mu_{\infty},
\label{chemical equilibrium}
\end{equation}
when ignoring the contribution of solvent molecules, other than the original PS formulation [6].

Meanwhile, the linear fitting to the above simulation results in the range of $5\times10^3\leq\Gamma\leq2\times10^6$ provides
\begin{equation}
\ln\Lambda=\Gamma+\mathrm{const.}
\label{Lambda coupling}
\end{equation}
(see eq. (58) in Ref. [15]).
Combining eqs. (\ref{mean surface}), (\ref{chemical equilibrium}) and (\ref{Lambda coupling}), we find
\begin{equation}
\mu_{\infty}=-2\Gamma,
\label{potential gamma}
\end{equation}
because there is obviously no proportionality of $\ln\rho(0)$ to $\Gamma$.
Furthermore, we have [3, 6. 7]
\begin{align}
\mu_{\infty}&=\frac{d}{d\sigma}\left(
\sigma u_{\infty}
\right)=\frac{3}{2}u_{\infty},
\label{chemical energy}
\end{align}
where use was made of the relation $du_{\infty}/d\sigma=u_{\infty}/(2\sigma)$ that is obtained from eq. (\ref{correlation}).
It follows from eqs. (\ref{potential gamma}) and (\ref{chemical energy}) that
\begin{equation}
u_{\infty}=-\frac{4}{3}\Gamma.
\label{correlation result}
\end{equation}
Namely, we obtain $\alpha=4/3$, which is larger than the previous prefactors of the OCPs: $\alpha_{2D}=1.1$ and $\alpha_{3D}=0.9$, with $\alpha_{2D}$ and $\alpha_{3D}$ denoting those of the 2D and 3D OCPs, respectively [16-25].

\subsection{Our aims}
The above PS two-phase model [3, 6, 7] has made use of the result $u_{\infty}=-\alpha_{2D}\Gamma\,(\alpha_{2D}=1.1)$, borrowing from the 2D OCP results, and yet there are no theories to provide $u_{\infty}$ solely for the counterion systems {\itshape from the first principle}.
Hence, our first aim is as follows:
\begin{itemize}
\item We develop a theoretical framework to calculate the ground-state energy $u_{\infty}$ per counterion. It was demonstrated in the OCPs that liquid-state theory [1] is relevant for this purpose. Therefore, we aim to incorporate the liquid-state theory into the framework for evaluating $u_{\infty}$ in strongly coupled counterion systems. 
\end{itemize}
Furthermore, eq. (\ref{correlation result}) indicates that an additional contribution to the difference between $\alpha=4/3$ and $\alpha_{2D}=1.1$ in $u_{\infty}$ needs to be explored. 
To this end, the thickness of the quasi-2D bound layer (region IV in Fig. 1) is investigated, which is our second aim:
\begin{itemize}
\item The theoretical issues to be addressed are twofold: (i) to demonstrate that the lowering of $u_{\infty}$ can be explained by applying the developed formulations to the quasi-2D bound state, and (ii) to verify that $\alpha=4/3$ can be reproduced only by introducing the thickness $\delta$ of molecular scale. 
\end{itemize}
It is noted that the developed theory can benefit from an analytic form of the direct correlation function that has been found to be relevant for the ground state [16-25].
We will use the soft mean spherical approximation (MSA) [20-23, 25] as described below.

\section{Results on the ground-state energy $u_{\infty}$}
\subsection{A lower-bound of the free energy $f_{\Gamma}$ per counterion in the ground state}
The Gibbs-Bogoliubov inequality [1, 25, 26] forms the basis of the lower-bound free energy $f_{\Gamma}$ per counterion.
The total free energy is expressed as $Nf_{\Gamma}$ using the total number $N$ of counterions, and the resulting free energy $Nf_{\Gamma}$ consists of energetic and entropic contributions that are functionals of the direct correlation function $c({\bf r})$ as well as the ground-state density $\rho_{\infty}({\bf r})$:
\begin{equation}
Nf_{\Gamma}=B[-c;\rho_{\infty}]+L[-c;\rho_{\infty}],
\label{lower free}
\end{equation}
where $B[-c;\rho_{\infty}]$ contributes to the lower-bound of $u_{\infty}$ and the entropic contribution is mainly of the logarithmic form due to the random phase approximation (RPA).
These functional forms of $B[-c;\rho_{\infty}]$ and $L[-c;\rho_{\infty}]$ are written as follows:
\begin{align}
B[-c;\rho_{\infty}]&=\frac{1}{2}\iint d{\bf r}d{\bf r}'\,
\left[\rho_{\infty}({\bf r})c({\bf r}-{\bf r}')\delta({\bf r}-{\bf r}')\right.\nonumber\\
&\qquad\left.
+\rho_{\infty}({\bf r})\rho_{\infty}({\bf r}')h({\bf r}-{\bf r}')\{c({\bf r}-{\bf r}')+v({\bf r}-{\bf r}')\}
\right],\label{functional1}\\
L[-c;\rho_{\infty}]&=\frac{1}{2}\ln\,\mathrm{det}\left\{
\delta({\bf r}-{\bf r}')-\rho_{\infty}({\bf r})c({\bf r}-{\bf r}')
\right\}-\int d{\bf r}\,\rho_{\infty}({\bf r}),
\label{functional2}
\end{align}
with $v({\bf r})\equiv\Gamma/|{\bf r}|$ and $h({\bf r})$ denoting the bare electrostatic interaction potential in the $k_BT$-unit and total correlation function, respectively.
It will be seen below that the above functionals are formulated for the quasi-2D system of counterions in the SC limit of $\Gamma\rightarrow\infty$.

\subsection{A general form of $u_{\infty}$}
The interaction energy density $u_{\Gamma}$ is obtained from $f_{\Gamma}$ as [1, 16, 17]
\begin{equation}
u_{\Gamma}=\Gamma\frac{\partial f_{\Gamma}}{\partial\Gamma}.
\label{u f}
\end{equation}
It is found from eqs. (\ref{lower free}) to (\ref{u f}) as well as the expression (\ref{correlation}) that the ground-state energy $u_{\infty}$ reads
\begin{align}
-\alpha&=\frac{u_{\infty}}{\Gamma}
=\lim_{\Gamma\rightarrow\infty}\frac{\partial f_{\Gamma}}{\partial\Gamma}=\lim_{\Gamma\rightarrow\infty}\frac{B[-c;\rho_{\infty}]}{N\Gamma},
\label{alpha}
\end{align}
where use has been made of the following relation:
\begin{equation}
\frac{\partial}{\partial\Gamma}
\left\{
c({\bf r})+v({\bf r})
\right\}=\frac{c({\bf r})+v({\bf r})}{\Gamma},
\end{equation}
which has a finite value even in the SC limit (see Appendix A for the details).
It is noted that the above result (\ref{alpha}) with $\rho_{\infty}$ replaced by uniform density smeared overall the system has successfully yielded $\alpha_{3D}$ of the 3D OCP [20-25], which is close to $0.9$, the prefactor of the Lieb-Narnhofer lower-bound [19]. 

\begin{figure}[hbtp]
\begin{center}
	\includegraphics[
	width=7 cm
]{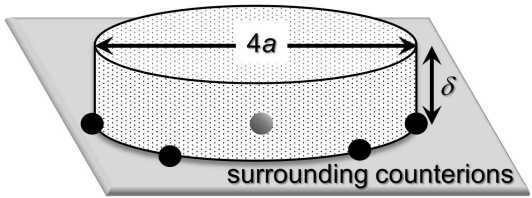}
\end{center}
\caption{Schematic of the cylindrical cell for a single counterion (a gray ball) bound on the planar surface.
The target counterion located along the central axis of the cylindrical cell is surrounded by adjacent counterions (black balls).
We can evaluate the ground-state energy per counterion in the SC limit by considering the integration ranges within the radius and height of $2a$ and $\delta$, respectively, in the corresponding cylindrical coordinate.
}
\end{figure}
\subsection{Cylindrical model for the description of the quasi-2D bound state}
In this study, we treat the bound counterions in the ground state, based on the combination of the soft MSA [20-23, 25] and cylindrical model depicted in Fig. 2:
the quasi-2D bound state in a single-counterion layer of thickness $\delta$ (or the region IV of Fig. 1) is defined by the soft MSA relations in the range of $0\leq z\leq \delta/a$, which read
\begin{flalign}
&c({\bf r}-{\bf r}')+v({\bf r}-{\bf r}')=0\quad\,|{\bf l}-{\bf l}'|>2,\label{soft msa1}\\
&h({\bf r}-{\bf r}')=-1\quad\quad\quad\qquad|{\bf l}-{\bf l}'|\leq 2,\label{soft msa2}
\end{flalign}
using the vector ${\bf l}=(x,y)$ on the $xy$-plane (i.e., ${\bf r}=({\bf l},z)$).
Here we should remember that the local electrical neutrality fixes the exclusion distance $2a$, as mentioned after eq. (1), though the global electrical neutrality of the bound layer (the region IV in Fig. 1) is violated even in the ground-state.

The cylindrical model corresponds to a coarse-grained description of the quasi-2D bound counterions in the proximity of the charged planar wall, or in the region IV of Fig. 1, as seen from the following interpretations of eq. (\ref{soft msa2}):
\begin{itemize}
\item {\itshape Longitudinal coarse-graining}.--- Equation (\ref{soft msa2}) ignores a degree of freedom in the z-axis direction, reflecting that the $z$-positions of adjacent counterions with local crystalline order vary independently within the single-counterion layer while maintaining the separation distance of $2a$ on the projective $xy$-plane.
\item {\itshape Transverse coarse-graining}.--- Equation (\ref{soft msa2}) also represents the coarse-grained single-counterion layer of thickness $\delta$ that allows no other counterions to enter from the top face of the cylinder occupied by a single counterion as shown in Fig. 2.
The quasi-2D bound state model (or the coarse-grained single-counterion layer model) validates that the mean-field treatment of adjacent counterions provides a constant density of $\rho_{\infty}({\bf 0})$ inside the cylinder, with the condition that the number $k$ of the neighboring counterions located on the circular edge of the cylinder in Fig. 2 should be approximately six (i.e., $k\approx 6$), considering the hexagonal packing of the 2D Wigner crystal.
\end{itemize}
Based on the above set of transverse and longitudinal views on the cylindrical model, we calculate the electrostatic interaction energy per quasi-2D bound counterion from focusing on the cylinder of Fig. 2 inside which the existence of adjacent counterions, interacting with the target counterion located along the central axis of the cylinder, is represented by the uniform density of $\rho_{\infty}({\bf 0})$.

\subsection{Evaluation of $u_{\infty}$ in the soft MSA of the cylindrical model}
It follows from eqs.  (\ref{soft msa1}) and (\ref{soft msa2}) that
\begin{align}
&\int d{\bf r}'\rho_{\infty}({\bf r}')
h({\bf r}-{\bf r}')\{c({\bf r}-{\bf r}')+v({\bf r}-{\bf r}')\}\nonumber\\
&\qquad=
-\int_{|{\bf r}-{\bf r}'|\leq 2} d{\bf r}'\rho_{\infty}({\bf r}')
\{c({\bf r}-{\bf r}')+v({\bf r}-{\bf r}')\}.
\label{msa integration}
\end{align}
Combination of eqs. (\ref{functional1}) and (\ref{msa integration}) yields an approximate form as follows:
\begin{flalign}
&\frac{B[-c;\rho_{\infty}]}{N}
=\frac{c({\bf 0})}{2}+\frac{\chi}{2}\nonumber\\
&\chi=-a^3\rho({\bf 0})\int_0^2dl\int_0^{\delta/a}dz\,(2\pi l)\,\{c({\bf r})+v({\bf r})\},
\label{bound general}
\end{flalign}
where we set a cylindrical cell for a single counterion located at the bottom center (see also Fig. 2) and introduce the cylindrical coordinate with the relation $r\equiv|{\bf r}|=\sqrt{l^2+z^2}$.
Not only eq. (\ref{bound general}) but also Fig. 2 further implies that the existence of surrounding counterions is taken into account by smearing the cylindrical cell with the density $a^3\rho({\bf 0})$, according to the PS two-phase model [3, 6, 7].

We can perform the integration in eq. (\ref{bound general}) using the analytic form of the direct correlation function in the soft MSA [20-23, 25] as detailed in Appendix A, providing
\begin{flalign}
-\alpha=\frac{u_{\infty}}{\Gamma}=\lim_{\Gamma\rightarrow\infty}\frac{B[-c;\rho_{\infty}]}{N\Gamma}
=-\frac{3}{5}-\frac{18}{35}\left(\frac{\delta}{\lambda}\right).
\label{u smsa}
\end{flalign}
Comparison between eqs. (\ref{correlation result}) and (\ref{u smsa}) leads to 
\begin{equation}
\delta=\frac{77}{54}\lambda,
\label{thickness result}
\end{equation}
which is our main result in this study;
the validity of $\delta$ will be investigated in the final section.

\section{Derivation scheme of the free energy functional given by eqs. (\ref{lower free}) to (\ref{functional2})}
This section aims to verify that $Nf_{\Gamma}$ corresponds to the lower-bound of the exact free energy $\Delta F[v]$ defined in Appendix C, where $\Delta$ represents the free energy difference from the vanishing base energy of bound state of counterions that are uniformly distributed on the planar surface (see also Appendix B). 
The Gibbs-Bogoliubov inequality [1, 25, 26] and the inhomogeneous RPA [26, 27] in a density-functional integral representation are key ingredient of the following formulations.

The Gibbs-Bogoliubov inequality [1, 25, 26] verifies that $\Delta F\{v\}$ has a lower-bound functional:
\begin{flalign}
&\Delta F[w]
+\frac{1}{2}\iint d{\bf r}d{\bf r}'\,\rho_{\infty}({\bf r})\rho_{\infty}({\bf r}')
\left[-w({\bf r}-{\bf r}')+v({\bf r}-{\bf r}')
\right]\nonumber\\
&\hphantom{\Delta F[w]+\frac{1}{2}\iint d{\bf r}d{\bf r}'\,\rho_{\infty}({\bf r})\rho_{\infty}({\bf r}')-w({\bf r}-{\bf r}')}
\leq\Delta F[v],
\label{GB}
\end{flalign}
where the lower-bound on the left hand side of the above inequality depends on an arbitrary interaction potential $w({\bf r})$, to be optimized, as well as on $h({\bf r})$, the actual correlation function instead of a reference function. 

Let $n({\bf r})$ be a density fluctuation around the ground-state distribution $\rho_{\infty}({\bf r})$. 
As derived in Appendix C, $\Delta F\{w\}$ can be expressed by the density-functional integral over the fluctuating $n$-field:
\begin{flalign}
e^{-\Delta F[w]}&=\int Dn\,e^{-{H}_w[n]},\nonumber\\
\mathcal{H}_w[n]&=\frac{1}{2}\iint d{\bf r}d{\bf r}'\,n({\bf r})n({\bf r}')w({\bf r}-{\bf r}')\nonumber\\
&\hphantom{\mathcal{H}_w[n]=\frac{1}{2}}
+\int d{\bf r}\left[
\frac{1}{2}\left\{
\frac{n^2({\bf r})}{\rho_{\infty}({\bf r})}-\rho_{\infty}({\bf r})w({\bf 0})
\right\}
-\rho_{\infty}({\bf r})
\right],
\label{df integral}
\end{flalign}
which reads the conventional RPA functional:
\begin{flalign}
\Delta F[w]
=-\frac{1}{2}\int d{\bf r}\,\rho_{\infty}({\bf r})w({\bf 0})
+L[w;\rho_{\infty}],
\label{rpa functional}
\end{flalign}
due to the Gaussian integration of eq. (\ref{df integral}) over the $n$-field.

The optimized lower-bound, or the maximum lower-bound, is determined by the stationary relation [26] as follows:
\begin{equation}
\left.
\frac{\delta(\Delta F\{w\})}{\delta w}\right|_{w=w^*}
=\frac{1}{2}h({\bf r}-{\bf r}').
\label{optimization}
\end{equation}
Equation (\ref{optimization}) with the use of the above RPA functional (\ref{rpa functional}) is reduced to the Ornstein-Zernike equation for inhomogeneous fluids (see Ref. [26] for the detailed derivation):
\begin{flalign}
&h({\bf r}-{\bf r}')=-w^*({\bf r}-{\bf r}')-\int d{\bf r}''h({\bf r}-{\bf r}'')\rho_{\infty}({\bf r}'')w^*({\bf r}''-{\bf r}'),&
\label{oz2}
\end{flalign}
thereby proving that minus the optimized potential $-w^*$ is identified with the direct correlation function:
\begin{equation}
w^*({\bf r}-{\bf r}')=-c({\bf r}-{\bf r}').
\label{w-c}
\end{equation}
Combining eqs. (\ref{GB}), (\ref{rpa functional}) and (\ref{w-c}), we have verified that the optimized lower-bound, which has been denoted by $Nf_{\Gamma}$ so far, is given by eqs. (\ref{lower free}) to (\ref{functional2}).

\section{Concluding remarks}
Returning to the 3D OCP, it is remembered that the Onsager smearing optimization [19-24] provided the correct ground-state energy or the Lieb--Narnhofer bound energy [19-24]. 
We have confirmed [20-24] that the lower bound approach presented here is equivalent to the Onsager smearing method in the ground state of the uniform 3D system.
 
The cylindrical coordinate given in eq. (\ref{bound general}) suggests the relationship between our treatment and the Onsager charge-smearing model, or the ionic sphere model [19-24]:
our results represented by eqs. (\ref{bound general}) and (\ref{u smsa}) imply the above correspondence between the lower bound approach and the Onsager model [19-24], extending to inhomogeneous quasi-2D systems.
In terms of the Onsager charge-smearing model [19-24], it can be seen that spherical charge smearing is adapted to the quasi-2D system by transforming into the cylindrical one.
Actually, the thickness $\delta$ given by eq. (\ref{thickness result}) in the mean-field approximation provides the following number $k$ of adjacent counterions:
\begin{equation}
k=(4\pi a^2\delta)\,\rho_{\infty}({\bf 0})\approx 5.7,
\label{number}
\end{equation}
which is close to six, the number of the hexagonal packing in the 2D Wigner crystal, consistently with either the Onsager charge-smearing model or the above view on the transverse coarse-graining (see section 3.3).

The remaining problem is to quantitatively assess the height $\delta$ of the cylindrical cell.
A quantitative evaluation of $\delta$ given by eq. (\ref{thickness result}) needs to be based on the site density, instead of the effective surface density reported in the literature; the latter is much smaller than the former due to the counterion effect.
We adopted a typical site density, $0.1/\mathrm{nm}^2\leq\sigma\leq0.5/\mathrm{nm}^2$, on glass and silica surfaces [28].
Thus, we can evaluate that $0.045\,\mathrm{nm}\leq\lambda\leq0.23\,\mathrm{nm}$ for the original GC length $\lambda$ when using $q=10$ and $l_B=0.7\,\mathrm{nm}$ in water solvent at room temperature.
It is found from eq. (\ref{thickness result}) that the cylindrical height $\delta$ is within the range of $0.065\,\mathrm{nm}\leq\delta\leq0.32\,\mathrm{nm}$.
The evaluated range of $\delta$ corroborates the supposition in the PS two-phase model [3, 6, 7] that $\delta$ is of the order of water molecular size (0.3 nm), as far as actually available systems are considered. 

Thus, we have achieved the present two aims: (i) we have incorporated the liquid-state theory into the theoretical framework developed for evaluating $u_{\infty}$, and also (ii) it has been demonstrated that $u_{\infty}=-(4/3)\Gamma$ implied by the recent simulation results can be obtained from considering a quasi-2D bound state with an adequate thickness $\delta$ of molecular scale. 

As a final remark, we should mention the overcharging phenomena [2-8]. 
Overcharging, or charge inversion, implies that the absolute value of opposite charges due to counterions accumulated on a macroion surface exceeds the bare charges that the macroion inherently carries. 
The highly favorable gain in the present correlation energy due to the bound counterions has been perceived as a promising candidate to elucidate the mechanism, particularly in low salt environments [2-8]; 
the PS two-phase model [3, 6, 7] is the pioneering work in this aspect. 
The developed method opens up the possibility of treating such types of complex phenomena more elaborately.

\section*{Declaration of competing interest}
The author declares that he has no known competing financial interests or personal relationships that could have appeared to influence the work reported in this paper.

\appendix
\section{Derivation of eq. (\ref{u smsa}) using the soft MSA form (\ref{msa dcf})}
We calculate the integration of $\chi$ given by eq. (\ref{bound general}), relying on an integration by parts formula as follows:
\begin{flalign}
\int_0^{\delta/a}dz\,f(l;z)&=
\left[zf(l;z)\right]_0^{\delta/a}-\int_0^{\delta/a}dz\,z\frac{df(l;z)}{dz}\nonumber\\
&=\left(\frac{\delta}{a}\right)f(l;\delta/a)-\int_{f(l;0)}^{f(l;\delta/a)}df\>z,
\label{by parts}
\end{flalign} 
where we set that $f(l;z)=2\pi l\,a^3\rho({\bf 0})\{c({\bf r})+v({\bf r})\}$ with $r=\sqrt{l^2+z^2}$.
Since we consider the thickness such that $\delta/a\ll 1$, eq. (\ref{by parts}) is reduced to
\begin{flalign}
\int_0^{\delta/a}dz\,f(l;z)&=\left(\frac{\delta}{a}\right)f(l;\delta/a)\nonumber\\
&=\left(2\pi l\delta a^2\right)\,\rho({\bf 0})\{c({\bf s})+v({\bf s})\},\nonumber\\
s&\equiv|{\bf s}|=\sqrt{l^2+(\delta/a)^2},
\label{reduced integral}
\end{flalign}
neglecting the $f$-integration range: $f(l;0)\approx f(l;\delta/a)$.
It follows that
\begin{flalign}
\frac{\chi}{2}&=-\frac{\delta a^2\rho({\bf 0})}{2}\int_{\delta/a}^{\sqrt{4+(\delta/a)^2}}ds\,(2\pi s)\,\{c({\bf s})+v({\bf s})\}\nonumber\\
&\approx
-\frac{\delta}{\lambda}\int_0^2ds\,(s)\,\{c({\bf s})+v({\bf s})\},
\label{chi reduced}
\end{flalign}
where use has been made of the relation $\pi a^2\rho({\bf 0})=1/\lambda$.

It has been found that the soft MSA of the direct correlation function provides the following form in the SC limit [20-23]: $-c({\bf r})=v({\bf r})=\Gamma/r$ for $r\equiv|{\bf r}|>2$, and
\begin{flalign}
\frac{-c({\bf r})}{\Gamma}
=\frac{6}{5}-\frac{1}{2}r^2+\frac{3}{16}r^3-\frac{1}{160}r^5
\label{msa dcf}
\end{flalign}
for $r\leq 2$. 
It follows from eqs. (\ref{chi reduced}) and (\ref{msa dcf}) that
\begin{flalign}
\frac{\chi}{2\Gamma}=\frac{\delta}{\lambda}\int_0^2ds\left(
\frac{6}{5}s-\frac{1}{2}s^3+\frac{3}{16}s^4-\frac{1}{160}s^6-1
\right)=-\frac{18}{35}\left(\frac{\delta}{\lambda}
\right),
\end{flalign} 
yielding eq. (\ref{u smsa}).

\section{Electrostatic interaction energies}
Let $\hat{\rho}({\bf r})$ be an instantaneous density of counterions located at ${\bf r}_i\,(i=1,\cdots ,N)$, where the counterion system is rescaled as ${\bf r}=(x,y,z)=(x_0/a,y_0/a,z_0/a)={\bf r}_0/a$.
The instantaneous density is expressed as
\begin{equation}
\hat{\rho}({\bf r})=a^3\sum_{i=1}^N\delta({\bf r}-{\bf r}_i), 
\end{equation}
the use of which the electrostatic interaction energy $\Delta U[v;{\hat{\rho}}]$ in the $k_BT$-unit is given by the sum of interaction energy differences, $\Delta U_{cc}[v;{\hat{\rho}}]$ and $\Delta U_{cm}[{\hat{\rho}}]$ that arise from counterion-counterion and counterion-wall interactions, respectively:
\begin{flalign}
&\Delta U[v;{\hat{\rho}}]=\Delta U_{cc}[v;{\hat{\rho}}]+\Delta U_{cm}[{\hat{\rho}}],\nonumber\\
&\Delta U_{cc}[v;{\hat{\rho}}]=\frac{\Gamma}{2}\iint
d{\bf r}d{\bf r}'v({\bf r}-{\bf r}')\nonumber\\
&\qquad\qquad\times\left[
\left\{\hat{\rho}({\bf r})-\rho_{\infty}({\bf r})\right\}
\left\{\hat{\rho}({\bf r}')-\rho_{\infty}({\bf r}')\right\}-\hat{\rho}({\bf r})\delta({\bf r}-{\bf r}')
\right],
\label{interaction cc}\\
&\Delta U_{cm}[\hat{\rho}]=2\Gamma\int d{\bf r}\>z\,\hat{\rho}({\bf r}),
\label{interaction cm}
\end{flalign}
where $\Delta$ represents the interaction energy difference from the vanishing base energy of bound state of counterions that are uniformly distributed on the planar surface. 

\section{Derivation of eqs. (\ref{df integral}) and (\ref{rpa functional})}
We start with the configurational representation of the free energy $\Delta F\{w\}$ with an interaction potential $w({\bf r}-{\bf r}')$ in the grand canonical system, which is expressed as
\begin{align}
e^{-\Delta F\{w\}}&=\mathrm{Tr}\,e^{-\Delta U[w;{\hat{\rho}}]},\nonumber\\
\mathrm{Tr}&\equiv
\sum_{N=0}^{\infty}\frac{e^{N\mu}}{N!}\int d{\bf r}\,_1\cdots\int d{\bf r}\,_N,
\label{f-start}
\end{align}
where $\Delta U[w;{\hat{\rho}}]$ is of the same form as eq. (\ref{interaction cc}) with $v$ replaced by $w$, and the chemical potential $\mu$ in the $k_BT$-unit has been introduced.
We relate the instantaneous density $\hat{\rho}({\bf r})$ to the density field $\rho_{\infty}({\bf r})+n({\bf r})$ with a fluctuating field $n({\bf r})$ added via the following identity [26]:
\begin{equation}
\int Dn\,\prod_{\{{\bf r}\}}\delta\left[\hat{\rho}({\bf r})-
\left\{\rho_{\infty}({\bf r})+n({\bf r})\right\}
\right]=1.
\label{identity}
\end{equation}
Plugging the Fourier transform of eq. (\ref{identity}) into eq. (\ref{f-start}), we have
\begin{align}
&e^{-\Delta F\{w\}}=\iint DnD\phi\,
e^{-\mathcal{H}_w[n;\phi]},\nonumber\\
&\mathcal{H}_w[n;\phi]=\frac{\Gamma}{2}\iint d{\bf r}d{\bf r}'\,w({\bf r}-{\bf r}')
\left\{n({\bf r})n({\bf r}')-\rho_{\infty}({\bf r})\delta({\bf r}-{\bf r}')\right\}\nonumber\\
&\qquad\qquad+\int d{\bf r}\left[i\phi({\bf r})\left\{\rho_{\infty}({\bf r})+n({\bf r})\right\}
-\rho_{\infty}({\bf r})e^{i\phi({\bf r})}
\right]
\label{n phi representation}
\end{align}
Expanding the exponential term of $\mathcal{H}_w[n;\phi]$ such that $\rho_{\infty}e^{i\phi}=\rho_{\infty}(1+i\phi-\phi^2/2)$, we can perform the Gaussian integration over the $\phi$-field.
Hence, eq. (\ref{n phi representation}) yields the quadratic density functional given in eq. (\ref{df integral}).
Furthermore, the Gaussian integration over the $n$-field leads to the result (\ref{rpa functional}) [26, 27].

\end{document}